# Anomalous Nernst effect in compensated ferrimagnetic Co$_x$Gd$_{1-x}$ films

Ruihao Liu[1,2], Li Cai[1,2], Teng Xu[1,2], Jiahao Liu[1,2], Yang Cheng[1,2] and Wanjun Jiang[1,2,a)]

[1]State Key Laboratory of Low-Dimensional Quantum Physics and Department of Physics, Tsinghua University, Beijing 100084, China
[2]Frontier Science Center for Quantum Information, Tsinghua University, Beijing 100084, China
[a)]Author to whom correspondence should be addressed: jiang_lab@tsinghua.edu.cn

**Abstract:** The anomalous Nernst effect (ANE) is one of the most intriguing thermoelectric phenomena which has attracted growing interest both for its underlying physics and potential applications. Typically, a large ANE response is observed in magnets with pronounced magnetizations or nontrivial Berry curvature. Here, we report a significant ANE signal in compensated ferrimagnetic Co$_x$Gd$_{1-x}$ alloy films, which exhibit vanishingly small magnetization. In particular, we found that the polarity of ANE signal is dominated by the magnetization orientation of the transition metal Co sublattices, rather than the net magnetization of Co$_x$Gd$_{1-x}$ films. This observation is not expected from the conventional understanding of ANE but is analogous to the anomalous Hall effect in compensated ferrimagnets. We attribute the origin of ANE and its Co-dominant property to the Co-dominant Berry curvature. Our work could trigger a more comprehensive understanding of ANE and may be useful for building energy-harvesting devices by employing ANE in compensated ferrimagnets.

Recent studies on thermoelectric effects have stimulated renewed understandings of their physical origins and also triggered increasing application opportunities[1–22]. The anomalous Nernst effect (ANE) is one of the representative examples[2-15]. When a longitudinal temperature gradient ($\nabla T$) is applied to a slab of magnetic material, the induced transverse electric field ($\boldsymbol{E}$) can be empirically formulated as[23,24]:

$$\boldsymbol{E} = Q_0(\boldsymbol{H_z} \times \nabla T) + Q_s(\mu_0 \boldsymbol{M_z} \times \nabla T) \qquad (1)$$

where $\mu_0$ is the vacuum permeability, $Q_0$ and $Q_s$ is the (ordinary/anomalous) Nernst coefficients, and $\boldsymbol{H_z}$, $\boldsymbol{M_z}$ are the applied perpendicular magnetic field, and the spontaneous net perpendicular magnetization, respectively. The two terms on the right side of Eq. (1) represent the ordinary Nernst effect (ONE) and ANE, respectively. Note that the contribution from ONE is typically smaller than that of ANE, which is thus often neglected[25]. From Eq. (1), it is clear that a large ANE is generally expected in ferromagnets with appreciable spontaneous magnetization[3,4,6,7,10,11,13]. Very recently, it is discovered that Eq. (1) becomes invalid in several antiferromagnets with no net magnetizations, where a large ANE could occur as a result of the nontrivial Berry curvature[9,15].

Compensated ferrimagnets (FIMs) are currently attracting considerable attention from the spintronics community for enabling the efficient spin-orbit torque switching[26-32] and the ultrafast domain wall motion[33–38]. Compensated FIMs exhibit minimal net magnetizations as a consequence of the antiparallel configuration of the involved magnetic sublattices[39,40]. Thus, it is not expected to observe pronounced ANE signals, as directly guided by Eq. (1). This intriguing aspect will be experimentally examined in the present study. Rare-earth-transition-metal (RE-TM) $Co_xGd_{1-x}$ amorphous films are typical compensated FIMs, in which the unequal magnetic moments of the Co and the Gd elements are antiferromagnetically coupled[39,40]. This results in a vanishingly small net magnetization ($M_z = M_{Co} - M_{Gd}$), where $M_{Co}$ and $M_{Gd}$ represent for the antiparallel magnetization of the Co and the Gd elements, as shown in Fig. 1(a). By changing the relative atomic composition ratio ($x$) between the Co and the Gd elements, or the external temperature ($T$), the contribution from each sublattice and the resultant magnetization could be effectively tuned. For example, crossing the critical composition compensation point ($x_{com} \approx 0.78$), or the temperature compensation point $T_{com}$, the dominant contribution to the net magnetization is either from the RE or TM elements, respectively[27,41–47].

In the present work, through utilizing an integrated on-chip method, we observe significant ANE signals in $Co_xGd_{1-x}$ films. Furthermore, the polarity of ANE signals in the RE-dominant sample is opposite to that of the TM-dominant sample. This observation is very similar to the behavior of the electrical counterpart of ANE, the anomalous Hall effect (AHE). This indicates that ANE could also be dominated by the magnetization of the TM sublattice, rather than the net magnetization. The origin of ANE and its TM-dominant property could be jointly attributed to the TM-dominant Berry curvature of the RE-TM compounds.

Magnetic multilayers of stacking order from the bottom to the top Ta(1)/Pt(3)/$Co_xGd_{1-x}$( $t_{CoGd}$ )/Ta(3) (numbers in parentheses are thicknesses in nanometers) are deposited on thermally oxidized silicon substrates using an AJA magnetron sputtering system (Orion 8). The base pressure of the main chamber is better than $1 \times 10^{-8}$ torr and the Ar pressure is 3.0 mtorr. The Co-dominant and Gd-dominant samples ($Co_{0.84}Gd_{0.16}$ and $Co_{0.74}Gd_{0.26}$, respectively) are synthesized through co-sputtering of the Co and Gd targets with a fixed sputtering power of Co (50 W, deposition rate 0.25 Å/s), while varying the power of Gd at 25 W (deposition rate 0.14 Å/s) and at 40 W (deposition rate 0.27 Å/s), respectively. Based on the deposition rate, the relative composition is determined (See Part 1 in Supplementary Materials). The thickness of the amorphous $Co_xGd_{1-x}$ layer is about 4.7 nm for $Co_{0.84}Gd_{0.16}$ and 6.2 nm for $Co_{0.74}Gd_{0.26}$. Through applying out-of-plane magnetic fields ( $H_z$ ), magnetic properties ( $M_z - H_z$ loop) are measured using a Superconducting Quantum Interference Device (SQUID, Quantum Design MPMS-3). Perpendicular magnetic anisotropy (PMA) is obtained in both samples, as shown in Fig. 1(b). It is displayed that the coercive field $H_c$ of these two samples are about 0.08 and 0.10 kOe, and the saturation net magnetization $M_S$ are about 260 emu/cc and 100 emu/cc, respectively. Magnitudes of $M_S$ in both samples are relatively small, which is consistent with the expectation of compensated ferrimagnetism[48].

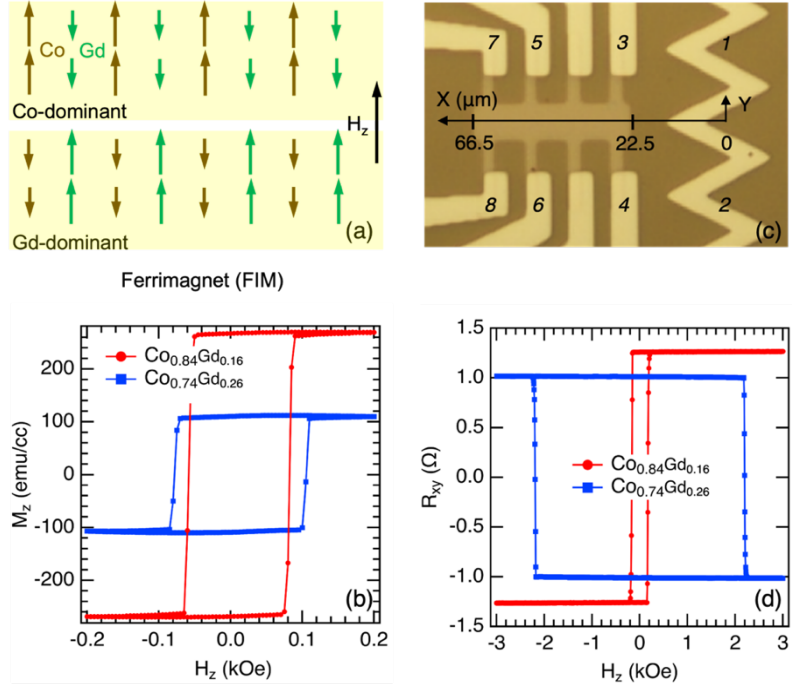

FIG. 1. (a): Schematic illustration of the collinear spin configurations in compensated ferrimagnet $Co_xGd_{1-x}$ with PMA. Showing in the upper panel is the spin configuration of the Co-dominant film, while the lower panel is the Gd-dominant film, respectively. The brown and green arrows represent the antiparallel and unequal magnetization of the Co and Gd elements, respectively. The black arrow represents the out-of-plane magnetic field. (b): The hysteresis loops ($M_z - H_z$) of the $Co_{0.84}Gd_{0.16}$ (red lines) and $Co_{0.74}Gd_{0.26}$ (blue lines) samples, respectively. (c): An optical image of the on-chip device that was made on a 100 nm thick $Si_3N_4$ insulating substrate. In this device, the AHE measurements are conducted by injecting an electric current from terminal 3 (or 4) to terminal 7 (or 8) and the AHE voltage can be obtained from terminals 5 and 6. The ANE measurements are conducted by injecting an electric current into the zigzag heater (from terminal 1 to terminal 2) and a transverse Nernst voltage can be obtained from terminals 3 and 4 (or 5 and 6 or 7 and 8, which are located at different positions from the on-chip heater). (d): The AHE loops ($R_{xy} - H_z$) of $Co_{0.84}Gd_{0.16}$ (red lines) and $Co_{0.74}Gd_{0.26}$ (blue lines), respectively.

These compensated FIMs are also deposited on the 100 nm thick $Si_3N_4$ membranes. Note that the $Si_3N_4$ membrane is electrically insulating, which, however, exhibits a relatively high in-plane thermal conductivity[49]. Devices are patterned by utilizing standard photolithography and lift-off techniques, as shown in Fig. 1(c). The AHE and ANE measurements are conducted on the same device by using a triple-axis superconducting magnet under the He atmosphere of 300 torr[13]. The AHE loops ($R_{xy} - H_z$) are displayed in Fig. 1(d) (See Supplementary Fig. S1 for the sign convention) and from which a PMA is further verified. Note that coercive fields ($H_c$) from the AHE measurements are larger than those from SQUID measurements (about 0.08 kOe and 0.10 kOe from SQUID and about 0.2 kOe and 2.2 kOe form the AHE measurement, respectively). This could be attributed to the additional edge-pinning effect that was introduced during the lithography process[50,51]. The AHE resistance $R_{xy}$ can be formulated from an empirical relation:

$$R_{xy} = R_0 \boldsymbol{H_z} + 4\pi R_s \boldsymbol{M_z} \qquad (2)$$

where $R_0$, $R_s$ are the ordinary and anomalous Hall coefficients. The contribution from the ordinary Hall effect (OHE) is negligible, as evident from the minimal change of $R_{xy}$ above saturation[52].

The SQUID magnetometry measures the net magnetization, which is parallel to the applied magnetic field above saturation[53]. Thus, magnetic hysteresis loops of these two samples should exhibit the same polarity. The AHE measurements in compensated FIMs, however, probe primarily the responses from the 3d shell of Co, due to the weak coupling between conduction electron spins with the inner 4f shell of the Gd magnetic moments[54-57]. Hence, the AHE responses are dominated by the magnetization configuration of the Co element in this FIM system[54-57]. This can be confirmed from the same polarity of magnetic hysteresis loops ($M_z - H_z$), as shown in Fig. 1(b), and the opposite polarity of AHE loops ($R_{xy} - H_z$) due to the opposite magnetization direction of the Co element, as shown in Fig. 1(d).

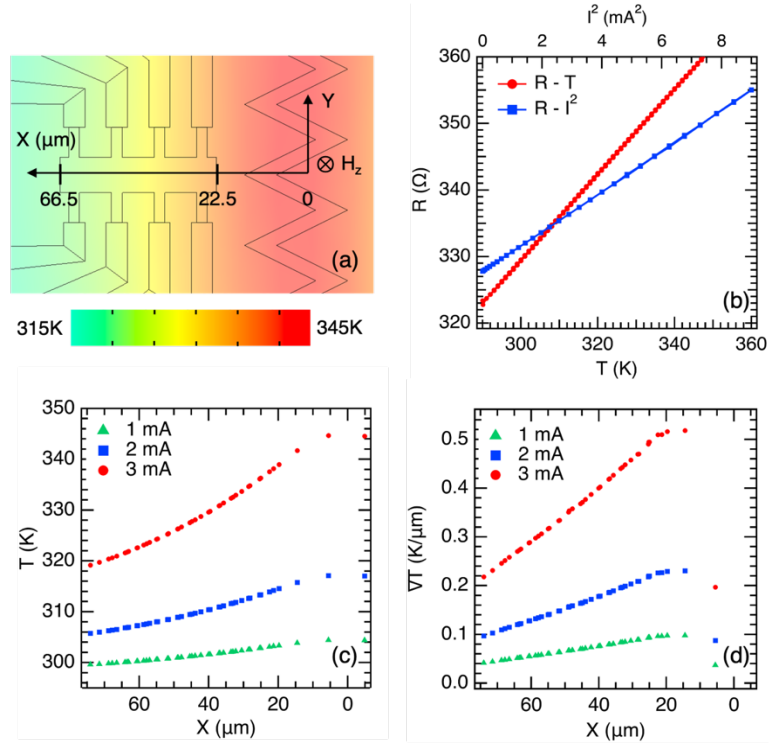

FIG. 2. (a): Heat distribution map of the device. This simulation is conducted by using the COMSOL software and incorporating the material-specific parameters. During the simulation, an electric current $I$ of 3.00 mA is injected into the heater. (b): Resistance of the heater as a function of temperature $T$ (red lines, lower horizontal axis) and as a function of $I^2$ (blue lines, upper horizontal axis). (c): Simulated temperature $T$ and (d): Simulated temperature gradient $\nabla T$ profile at different positions when injecting $I$ of 1.00 mA (green dots), 2.00 mA (blue dots), and 3.00 mA (red dots), respectively.

As shown in Fig. 1(c), the ANE measurements are conducted by first injecting an electric current $I$ into the on-chip heater. To identify whether the dissipation of Joule heating from the heater generates a temperature gradient along the $X$ direction, a finite element simulation using the COMSOL software is performed. By using $I = 3.00$ mA, the simulated heat distribution map is shown in Fig. 2(a). By separately measuring the temperature-dependent resistance changes of the heater, and the resistance changes as a function of $I^2$ at (nominal) room temperature, the temperature increase due to Joule heating can thus be estimated, as shown in Fig. 2(b). The evolution of the temperature $T$ and temperature gradient $\nabla T$ along the $X$ axis, is simulated and shown in Figs. 2(c) and 2(d), respectively. It is demonstrated that, within the range of $X > 20$ μm, both the temperature and temperature gradient decrease monotonously as the distance from the heater increases. Similarly, a larger electrical current corresponds to a larger temperature gradient at a fixed position. These simulation results suggest that an appreciable and evolving temperature gradient $\nabla T$ could be established in the current device geometry, which ensures the ANE measurements. As a result, a transverse Nernst voltage can be observed from three different channels (3 μm ($X$) × 27 μm ($Y$), at $X$ of 26.5 μm, 50.5 μm, and 62.5 μm), respectively. During the ANE measurements, the second harmonic signal is detected, since the current-induced Joule heating is proportional to $I^2$.

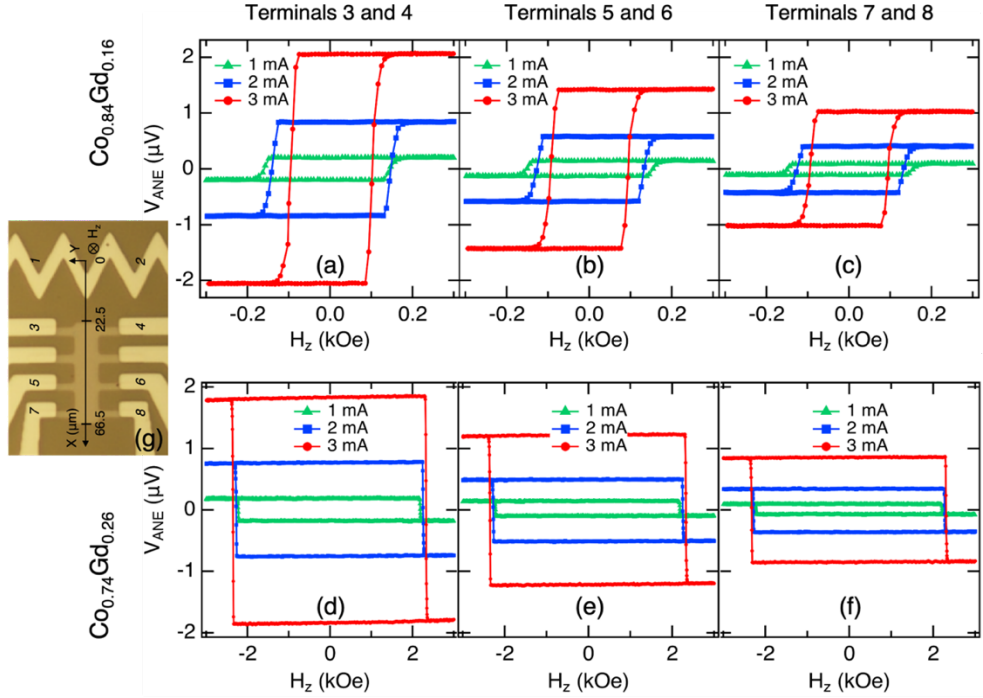

FIG. 3. The ANE voltages ($V_{ANE}$) as a function of $H_z$ when injecting a current of 1.00 mA (green lines, the estimated temperature gradients: 0.09 K/μm, 0.07 K/μm and 0.05 K/μm from the left to the right column), 2.00 mA (blue lines, temperature gradients: 0.21 K/μm, 0.15 K/μm and 0.12 K/μm) and 3.00 mA (red lines, temperature gradients: 0.48 K/μm, 0.34 K/μm and 0.28 K/μm) into the heater, measured from three different channels (three columns), in $Co_{0.84}Gd_{0.16}$ (upper row) and $Co_{0.74}Gd_{0.26}$ (lower row), respectively. (g): The optical image of the on-chip ANE device.

The $V_{ANE}$ *vs.* $H_z$ loops are shown in Figs. 3(a)-(c) for Co$_{0.84}$Gd$_{0.16}$ and Figs. 3(d)-(f) for Co$_{0.74}$Gd$_{0.26}$, respectively (See Supplementary Fig. S1 for the sign convention). More raw data are presented in Fig. S2 through injecting various currents into the heater. Similar to AHE, slopes of the ANE loops in the saturation regime only exhibit minor variations, as expected from the negligible contribution of the ONE signal. Its contribution is thus neglected in the following discussion. The observed ANE loops in the Co$_{0.84}$Gd$_{0.16}$ and Co$_{0.74}$Gd$_{0.26}$ films exhibit an opposite sign, which is analogous to the AHE behavior. Furthermore, the coercive fields $H_c$ are nearly the same from the AHE and ANE measurements. These data demonstrate that ANE acts similarly to AHE, indicating that the ANE response could also be dominated by the magnetization of the TM element (Co) in RE-TM (Co$_x$Gd$_{1-x}$) compensated FIMs.

Phenomenologically, the analogy between the AHE and ANE behaviors could be explained as follows: the dominant contribution to thermal transport is from the conduction electrons in the present metallic system[2-5,22], rather than bosonic phonons or magnons, as indicated by the Wiedemann-Franz Law[58,59], and the elcetrons driven by the tempareture gradient directly leads to ANE. This is similar to the electrical transport phenomena, such as AHE[52]. Furthermore, several early works have already suggested that these two effects are intrinsically connected, as implied by the Mott relation[2-5,22]. Thus, due to similar reasons elucidating AHE in RE-TM materials, the TM-dominant property of ANE could be qualitatively understood.

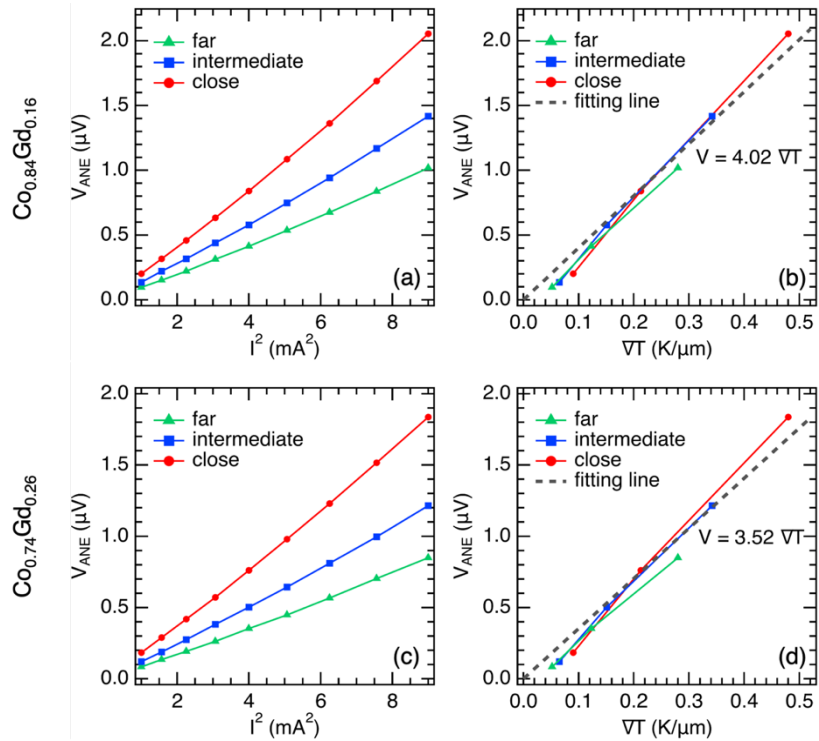

FIG. 4. The obtained ANE voltages $V_{ANE}$ as a function of injected current $I^2$ and the simulated temperature gradient $\nabla T$, at different channel positions close to the heater (red lines), intermediate (blue lines), and far from the heater (green lines), for Co$_{0.84}$Gd$_{0.16}$: (a)/(b) and Co$_{0.74}$Gd$_{0.76}$: (c)/(d). Linear fitting lines (gray lines) and corresponding functions are also shown in (b) and (d).

As suggested by Eq. (1), the linear dependence of ANE on $\nabla T$ is testified by the evolution of $V_{ANE}$ as a function of $I^2$, and the extracted $\nabla T$ from the simulations, as shown in Figs. 4(a)-4(b) for $Co_{0.84}Gd_{0.16}$ and Figs. 4(c)-4(d) for $Co_{0.74}Gd_{0.26}$, respectively. Furthermore, by linearly fitting the $\nabla T$ dependence of $V_{ANE}$, as shown in Figs. 4(b) and 4(d), the thermopower of ANE can be estimated as $S_{yx} \approx 0.15 \mu V/K$ for $Co_{0.84}Gd_{0.16}$ and $S_{yx} \approx -0.13 \mu V/K$ for $Co_{0.74}Gd_{0.26}$, respectively.

TABLE I. The estimated tangent of the anomalous Hall angle ($tan\theta_{AHE}$), thermopower of the SE/ANE ($S_{xx}/S_{yx}$), the first/second term of the linear response theory ($S_I/S_{II}$) and the transverse thermoelectric conductivity ($\alpha_{yx}$) of $Co_{0.84}Gd_{0.16}$ and $Co_{0.74}Gd_{0.26}$.

| Sample | $tan\theta_{AHE}$ | $S_{xx}$[a] ($\mu V\,K^{-1}$) | $S_{yx}$ ($\mu V\,K^{-1}$) | $S_I$ ($\mu V\,K^{-1}$) | $S_{II}$ ($\mu V\,K^{-1}$) | $\alpha_{yx}$ ($A\,m^{-1}\,K^{-1}$) |
|---|---|---|---|---|---|---|
| $Co_{0.84}Gd_{0.16}$ | -0.003 | -8 | 0.15 | 0.126 | 0.024 | 0.071 |
| $Co_{0.74}Gd_{0.26}$ | 0.002 | -4 | -0.13 | -0.122 | -0.008 | -0.044 |

[a] Data are collected from Ref. 17.

To study the origin and quantitatively demonstrate the Co-dominant property of ANE in $Co_xGd_{1-x}$ films, we utilize the linear response theory[60-62]:

$$S_{yx} = \rho_{xx}\alpha_{yx} + \rho_{yx}\alpha_{xx}, (3)$$

where $\rho_{xx}/\rho_{yx}$ and $\alpha_{xx}/\alpha_{yx}$ are the longitudinal/transverse electrical resistivity and longitudinal/transverse thermoelectric conductivity, respectively. We denote the first and the second term on the right-hand side of Eq. (3) as $S_I = \rho_{xx}\alpha_{yx}$ and $S_{II} = \rho_{yx}\alpha_{xx}$. Note that $S_{II}$ can be converted into $S_{II} = tan\theta_{AHE}\,S_{xx}$, where $\theta_{AHE} = arctan\,(\rho_{yx}/\rho_{xx})$ is the anomalous Hall angle, and $S_{xx}$ is the thermopower of the Seebeck effect (SE). Here, $S_I$ represents the contribution from the intrinsic ANE to the total ANE signal, while $S_{II}$ appears as a consequence of the product of the SE and the AHE.

These parameters are summarized in Table I. Since $|S_I| \gg |S_{II}|$, $S_I$ (or $\alpha_{yx}$) can be regarded as the dominant contribution to the $S_{yx}$ (ANE). As $\alpha_{yx}$ stems primarily from the Berry curvature of the electron bands near the Fermi level[11,62], the origin of ANE could thus be interpreted. Furthermore, the sign of $\alpha_{yx}$ is opposite in the two samples, which plays the dominant role in the different ANE polarities. As the electrons near the Fermi level are dominated by the 3d band of Co sublattices[56] and the spin orientations of which are opposite in the Co/Gd-dominant samples, this could lead to distinctively different contribution from Berry curvatures, the sign change of $\alpha_{yx}$ and consequently the sign change of ANE could occur. This thus justifies the TM-dominant contribution of ANE in RE-TM FIMs.

In conclusion, we have synthesized Co-dominant ($Co_{0.84}Gd_{0.16}$) and Gd-dominant ($Co_{0.74}Gd_{0.26}$) compensated ferrimagnetic films with perpendicular magnetic anisotropy, in which the magnetic properties and transport properties are systematically examined. Appreciable ANE signals are demonstrated in compensated ferrimagnets. Furthermore, it is found that the polarities of ANE are opposite in the two samples, which is analogous to the AHE behavior. These observations demonstrate that the ANE responses could also be intrinsically connected with the magnetism of the transition metal element, rather than the net magnetization in the RE-TM compensated ferrimagnets. The origin and the transition metal-dominant property of ANE could be understood by invoking the Berry curvature of the TM component. More importantly, it seems quite promising to make transverse thermoelectric devices, thermal piles for example, based on the opposite signs of ANE in the compensated ferrimagnets, which could output large thermoelectric voltages while exhibiting moderate stray fields.

See the Supplementary Materials for the composition characterizations of the $Co_xGd_{1-x}$ films, sign conventions used in AHE and ANE measurements, raw data of ANE loops, X-ray diffraction measurements for evidence of the amorphous $Co_xGd_{1-x}$ films, and the schematic device structure and parameters used in the COMSOL simulations.

This work was supported by the general program of NSFC (Grant Nos. 52271181, 51831005), the National Natural Science Foundation of China (NSFC) under the distinguished Young Scholar program (Grant No. 12225409), the Beijing Natural Science Foundation (Grant No. Z190009), the Tsinghua University Initiative Scientific Research Program and the Beijing Advanced Innovation Center for Future Chip (ICFC).

**Data Availability**

The data that support the findings of this study are available from the corresponding author upon reasonable request.

**References:**

[1] K. Uchida, S. Takahashi, K. Harii, J. Ieda, W. Koshibae, K. Ando, S. Maekawa, and E. Saitoh, Nature **455**, 778 (2008).
[2] D. Xiao, Y. Yao, Z. Fang, and Q. Niu, Phys. Rev. Lett. **97**, 026603 (2006).
[3] T. Miyasato, N. Abe, T. Fujii, A. Asamitsu, S. Onoda, Y. Onose, N. Nagaosa, and Y. Tokura, Phys. Rev. Lett. **99**, 086602 (2007).
[4] Y. Pu, D. Chiba, F. Matsukura, H. Ohno, and J. Shi, Phys. Rev. Lett. **101**, 117208 (2008).
[5] S. Onoda, N. Sugimoto, and N. Nagaosa, Phys. Rev. B **77**, 165103 (2008).

[6] M. Mizuguchi, S. Ohata, K. Uchida, E. Saitoh, and K. Takanashi, Appl. Phys. Express **5**, 093002 (2012).

[7] Y. Sakuraba, K. Hasegawa, M. Mizuguchi, T. Kubota, S. Mizukami, T. Miyazaki, and K. Takanashi, Appl. Phys. Express **6**, 033003 (2013).

[8] K. Hasegawa, M. Mizuguchi, Y. Sakuraba, T. Kamada, T. Kojima, T. Kubota, S. Mizukami, T. Miyazaki, and K. Takanashi, Appl. Phys. Lett. **106**, 252405 (2015).

[9] M. Ikhlas, T. Tomita, T. Koretsune, M.-T. Suzuki, D. Nishio-Hamane, R. Arita, Y. Otani, and S. Nakatsuji, Nat. Phys. **13**, 1085 (2017).

[10] T.C. Chuang, P.L. Su, P.H. Wu, and S.Y. Huang, Phys. Rev. B **96**, 174406 (2017).

[11] A. Sakai, Y.P. Mizuta, A.A. Nugroho, R. Sihombing, T. Koretsune, M.-T. Suzuki, N. Takemori, R. Ishii, D. Nishio-Hamane, R. Arita, P. Goswami, and S. Nakatsuji, Nat. Phys. **14**, 1119 (2018).

[12] H. Reichlova, R. Schlitz, S. Beckert, P. Swekis, A. Markou, Y.-C. Chen, S. Fabretti, G.H. Park, A. Niemann, S. Sudheendra, A. Thomas, K. Nielsch, C. Felser, and S.T.B. Goennenwein, Appl. Phys. Lett. **113**, 212405 (2018).

[13] J. Hu, T. Butler, M.A. Cabero Z., H. Wang, B. Wei, S. Tu, C. Guo, C. Wan, X. Han, S. Liu, W. Zhao, J.-P. Ansermet, S. Granville, and H. Yu, Appl. Phys. Lett. **117**, 062405 (2020).

[14] T. Chen, T. Tomita, S. Minami, M. Fu, T. Koretsune, M. Kitatani, I. Muhammad, D. Nishio-Hamane, R. Ishii, F. Ishii, R. Arita, and S. Nakatsuji, Nat. Commun. **12**, 572 (2021).

[15] Y. Pan, C. Le, B. He, S.J. Watzman, M. Yao, J. Gooth, J.P. Heremans, Y. Sun, and C. Felser, Nat. Mater. **21**, 203 (2022).

[16] S. Meyer, Y.-T. Chen, S. Wimmer, M. Althammer, T. Wimmer, R. Schlitz, S. Geprägs, H. Huebl, D. Ködderitzsch, H. Ebert, G.E.W. Bauer, R. Gross, and S.T.B. Goennenwein, Nat. Mater. **16**, 977 (2017).

[17] T. Seki, A. Miura, K. Uchida, T. Kubota, and K. Takanashi, Appl. Phys. Express **12**, 023006 (2019).

[18] V. Popescu, P. Kratzer, P. Entel, C. Heiliger, M. Czerner, K. Tauber, F. Töpler, C. Herschbach, D.V. Fedorov, M. Gradhand, I. Mertig, R. Kováčik, P. Mavropoulos, D. Wortmann, S. Blügel, F. Freimuth, Y. Mokrousov, S. Wimmer, D. Ködderitzsch, M. Seemann, K. Chadova, and H. Ebert, J. Phys. Appl. Phys. **52**, 073001 (2019).

[19] W. Zhou, K. Yamamoto, A. Miura, R. Iguchi, Y. Miura, K. Uchida, and Y. Sakuraba, Nat. Mater. **20**, 463 (2021).

[20] G.E.W. Bauer, E. Saitoh, and B.J. van Wees, Nat. Mater. **11**, 391 (2012).

[21] S.R. Boona, R.C. Myers, and J.P. Heremans, Energy Environ. Sci. **7**, 885 (2014).

[22] K. Vandaele, S.J. Watzman, B. Flebus, A. Prakash, Y. Zheng, S.R. Boona, and J.P. Heremans, Mater. Today Phys. **1**, 39 (2017).

[23] W. Nernst, Ann. Phys. **267**, 760 (1887).

[24] M. Mizuguchi and S. Nakatsuji, Sci. Technol. Adv. Mater. **20**, 262 (2019).

[25] K. Uchida, Proc. Jpn. Acad. Ser. B **97**, 69 (2021).

[26] J. Finley and L. Liu, Phys. Rev. Appl. **6**, 054001 (2016).

[27] R. Mishra, J. Yu, X. Qiu, M. Motapothula, T. Venkatesan, and H. Yang, Phys. Rev. Lett. **118**, 167201 (2017).

[28] A. Manchon, J. Železný, I.M. Miron, T. Jungwirth, J. Sinova, A. Thiaville, K. Garello, and P. Gambardella, Rev. Mod. Phys. **91**, 035004 (2019).

[29] H. Wu, Y. Xu, P. Deng, Q. Pan, S.A. Razavi, K. Wong, L. Huang, B. Dai, Q. Shao, G. Yu, X. Han, J. Rojas-Sánchez, S. Mangin, and K.L. Wang, Adv. Mater. **31**, 1901681 (2019).

[30] K. Cai, Z. Zhu, J.M. Lee, R. Mishra, L. Ren, S.D. Pollard, P. He, G. Liang, K.L. Teo, and H. Yang, Nat. Electron. **3**, 37 (2020).

[31] C. Song, R. Zhang, L. Liao, Y. Zhou, X. Zhou, R. Chen, Y. You, X. Chen, and F. Pan, Prog. Mater. Sci. **118**, 100761 (2021).

[32] Q. Shao, P. Li, L. Liu, H. Yang, S. Fukami, A. Razavi, H. Wu, K. Wang, F. Freimuth, Y. Mokrousov, M.D. Stiles, S. Emori, A. Hoffmann, J. Åkerman, K. Roy, J.-P. Wang, S.-H. Yang, K. Garello, and W. Zhang, IEEE Trans. Magn. **57**, 1 (2021).

[33] K.-J. Kim, S.K. Kim, Y. Hirata, S.-H. Oh, T. Tono, D.-H. Kim, T. Okuno, W.S. Ham, S. Kim, G. Go, Y. Tserkovnyak, A. Tsukamoto, T. Moriyama, K.-J. Lee, and T. Ono, Nat. Mater. **16**, 1187 (2017).

[34] S.A. Siddiqui, J. Han, J.T. Finley, C.A. Ross, and L. Liu, Phys. Rev. Lett. **121**, 057701 (2018).

[35] L. Caretta, M. Mann, F. Büttner, K. Ueda, B. Pfau, C.M. Günther, P. Hessing, A. Churikova, C. Klose, M. Schneider, D. Engel, C. Marcus, D. Bono, K. Bagschik, S. Eisebitt, and G.S.D. Beach, Nat. Nanotechnol. **13**, 1154 (2018).

[36] S. Vélez, J. Schaab, M.S. Wörnle, M. Müller, E. Gradauskaite, P. Welter, C. Gutgsell, C. Nistor, C.L. Degen, M. Trassin, M. Fiebig, and P. Gambardella, Nat. Commun. **10**, 4750 (2019).

[37] L. Caretta, S.-H. Oh, T. Fakhrul, D.-K. Lee, B.H. Lee, S.K. Kim, C.A. Ross, K.-J. Lee, and G.S.D. Beach, Science **370**, 1438 (2020).

[38] H.-A. Zhou, T. Xu, H. Bai, and W. Jiang, J. Phys. Soc. Jpn. **90**, 081006 (2021).

[39] J. Finley and L. Liu, Appl. Phys. Lett. **116**, 110501 (2020).

[40] S.K. Kim, G.S.D. Beach, K.-J. Lee, T. Ono, T. Rasing, and H. Yang, Nat. Mater. **21**, 24 (2022).

[41] S. Demirtas, R.E. Camley, and A.R. Koymen, Appl. Phys. Lett. **87**, 202111 (2005).

[42] C. Kaiser, A.F. Panchula, and S.S.P. Parkin, Phys. Rev. Lett. **95**, 047202 (2005).

[43] X. Jiang, L. Gao, J.Z. Sun, and S.S.P. Parkin, Phys. Rev. Lett. **97**, 217202 (2006).

[44] W. Zhou, T. Seki, T. Kubota, G.E.W. Bauer, and K. Takanashi, Phys. Rev. Mater. **2**, 094404 (2018).

[45] C.E. Patrick and J.B. Staunton, Phys. Rev. B **97**, 224415 (2018).

[46] T. Fu, S. Li, X. Feng, Y. Cui, J. Yao, B. Wang, J. Cao, Z. Shi, D. Xue, and X. Fan, Phys. Rev. B **103**, 064432 (2021).

[47] G. Sala, C.-H. Lambert, S. Finizio, V. Raposo, V. Krizakova, G. Krishnaswamy, M. Weigand, J. Raabe, M.D. Rossell, E. Martinez, and P. Gambardella, Nat. Mater. **21**, 640 (2022).

[48] J.M.D. Coey, *Magnetism and Magnetic Materials* (Cambridge University Press, 2010).

[49] Z. Wang, M. Guo, H.-A. Zhou, L. Zhao, T. Xu, R. Tomasello, H. Bai, Y. Dong, S.-G. Je, W. Chao, H.-S. Han, S. Lee, K.-S. Lee, Y. Yao, W. Han, C. Song, H. Wu, M.

Carpentieri, G. Finocchio, M.-Y. Im, S.-Z. Lin, and W. Jiang, Nat. Electron. **3**, 672 (2020).

[50] F. Dumestre, B. Chaudret, C. Amiens, M.-C. Fromen, M.-J. Casanove, P. Renaud, and P. Zurcher, Angew. Chem. Int. Ed. **41**, 4286 (2002).

[51] F. Ott, T. Maurer, G. Chaboussant, Y. Soumare, J.-Y. Piquemal, and G. Viau, J. Appl. Phys. **105**, 013915 (2009).

[52] N. Nagaosa, J. Sinova, S. Onoda, A.H. MacDonald, and N.P. Ong, Rev. Mod. Phys. **82**, 1539 (2010).

[53] R. L. Fagaly, Rev. Sci. Instrum. **77**, 101101 (2006).

[54] T. Shirakawa, Y. Nakajima, K. Okamoto, S. Matsushita, and Y. Sakurai, AIP Conf. Proc. **34**, 349 (1976).

[55] Y. Mimura, N. Imamura, and Y. Kushiro, J. Appl. Phys. **47**, 3371 (1976).

[56] H. Tanaka, S. Takayama, and T. Fujiwara, Phys. Rev. B **46**, 7390 (1992).

[57] T. Xu, H.-A. Zhou, Y. Dong, Q. Zhang, M. Che, L. Liu, Z. Wu, Z. Guan, L. Yang, and W. Jiang, Phys. Rev. Appl. **16**, 044056 (2021).

[58] G.V. Chester and A. Thellung, Proc. Phys. Soc. **77**, 1005 (1961).

[59] R.J. Anderson, J. Appl. Phys. **67**, 6914 (1990).

[60] K. Sumida, Y. Sakuraba, K. Masuda, T. Kono, M. Kakoki, K. Goto, W. Zhou, K. Miyamoto, Y. Miura, T. Okuda, and A. Kimura, Commun. Mater. **1**, 89 (2020).

[61] Y. Sakuraba, K. Hyodo, A. Sakuma, and S. Mitani, Phys. Rev. B **101**, 134407 (2020).

[62] K. Uchida, W. Zhou, and Y. Sakuraba, Appl. Phys. Lett. **118**, 140504 (2021).